# Injection Locking of Optoelectronic Oscillators with Large Delay

Mehedi Hasan, Abhijit Banerjee, and Trevor James Hall

The optoelectronic oscillator is a delay line oscillator that leverages optical fiber technology to realize the large delay required for low phase noise systems. Spurious sidemodes are an artifact of the delay line oscillator, yet treatments of injection locking of optoelectronic oscillators have relied on the application of classical injection locking theory valid only for single mode oscillators. The large delay contributed by the optical fiber delay line is accounted for by classical theory only in part through the quality factor Q that captures the round-trip group delay in a neighborhood of the oscillation frequency. This paper presents a new formulation of time delay oscillators subject to injection that describes all the essential features of their dynamics and phase noise. The common assumptions of a single mode oscillator and weak injection are removed. This is important to correctly predict the locking range, the suppression of sidemodes and the phase noise spectrum. The findings of the analysis are validated by experimental measurements provided by an optoelectronic oscillator under injection by an external source.

## I. INTRODUCTION

The random fluctuations of the phase of an oscillator limit the precision of time and frequency measurements. Consequently, the noise and long-term stability of the system oscillator / clock is of major importance in applications such as optical and wireless communications, high-speed digital electronics, radar, and astronomy [1]-[7]. Delay line oscillators based on photonic components, offer the potential for realization of phase noise levels up to three orders of magnitude lower than achievable by conventional microwave sources. The optoelectronic oscillator (OEO) introduced by Steve Yao and Lute Maleki in 1996 [8] leverages optical fiber technology to realize the large delay required for low phase noise systems whilst simultaneously achieving insertion loss levels that can be compensated by available microwave and photonic amplification technologies. Multimode operation is an artifact of the delay line oscillator which introduces spurious resonances within the phase noise spectrum and can introduce instability within a phased lock loop (PLL) [9]. A variety of architectures and methods have been proposed in pursuit of a compact low phase noise OEO with suppressed spurious resonances. Proposed solutions include external-, self- or cross- injection locking of single-loop, multi-loop, or multi-oscillator architectures [10]-[17].

The theory of injection locking and pulling of oscillators was pioneered by Adler [18] with notable contributions by Paciorek [19], Armand [20], and Couch [21] while the model of oscillator phase noise introduced by Leeson [22]-[23] is ubiquitous on account of its utility. Prior theoretical descriptions of injection locking of optoelectronic oscillators have generally relied on the application of classical injection locking theory which is strictly valid only for single mode oscillators. The large delay contributed by the optical fiber delay line is accounted for only in part through the quality factor Q that captures the loop group delay in a neighborhood of the oscillation frequency [24].

This paper presents a new formulation of time delay oscillators subject to injection that describes all the essential features of their dynamics and phase noise. The common assumptions of a single mode oscillator and weak injection are removed. This is important to correctly predict the effect of injection locking on the spurious resonances in the phase noise spectrum corresponding to the sidemodes of a time delay oscillator. A single loop OEO injection locked by an external source is used as an example to elucidate the fundamental theory but it is straightforward to apply the formulation to time delay oscillator topologies of greater complexity. The dynamics of the oscillator under injection and analytic results on the locking range and phase noise spectrum predicted by the theory are validated by experimental results provided by a laboratory prototype OEO.

This paper is organized as follows. In Section II an envelope model of a free single loop OEO is formulated. Application of the Lesson postulate to the envelope model leads to a linear phase-only model of the oscillator which is used to predict the free oscillator phase noise in terms of the spectral density of intra-loop phase fluctuations. In Section III the formulation is extended to a nonlinear delay differential equation phase-only model that describes a single loop OEO subject to injection. In Section IV a comparison is made to the classical theory of injection locking. Section V considers perturbations about equilibrium solutions leading to a simple accurate linear phase-only model which is used to predict the phase noise spectrum of the injection locked OEO. Experimental observations are presented in Section VI to validate the predictions of the theory. Section VII concludes the paper with a summary and discussion of the main findings.

## II. FREE SINGLE LOOP OEO

An OEO is a time delay oscillator. It consists of the concatenation of a RF-photonic link and an RF amplifier chain configured into a loop as shown in Fig. 1. The purpose of the RF-photonic link is to provide the time delay of the oscillator. The purpose of the RF chain is to provide the sustaining amplifier of the oscillator.

The RF photonic link consists of a laser, a Mach-Zehnder modulator (MZM) and a photo-receiver. The optical carrier provided by the laser is intensity modulated by the RF-input applied to the MZM. This is followed by transmission through a long optical fiber to a photo-receiver that recovers the RF modulation.

The RF chain consists of an RF amplifier followed by an RF band-pass filter (BPF). At sufficiently small input signal levels the RF amplifier provides linear gain but at higher levels hard clipping generates harmonics which are dissipated by the RF BPF[1]. If the spectrum of a signal before clipping is narrowband the spectra of the most significant harmonics will not overlap the fundamental. The carrier generated by an oscillator has a spectrum that is broadened only by low-frequency phase noise modulation. It is therefore reasonable to assume that the RF BPF is sufficiently selective to substantially suppress the harmonics of the oscillation generated by amplifier nonlinearity yet has a sufficient passband to substantially pass unaltered low frequency phase modulation of the fundamental. The magnitude of the bandpass-filtered output versus the magnitude of the input of the amplifier will saturate at input levels beyond the linear region and the linearized gain will approach zero substantially suppressing any fluctuations in magnitude of the output. The RF amplifier-filter chain operated in saturation thereby provides a substantially constant magnitude output. The dynamical system describing a single loop OEO consequently takes the form:

$$u = e^{i\phi}\left(h \otimes \left(K(D_{\tau_D}u)\right)\right) \tag{1}$$

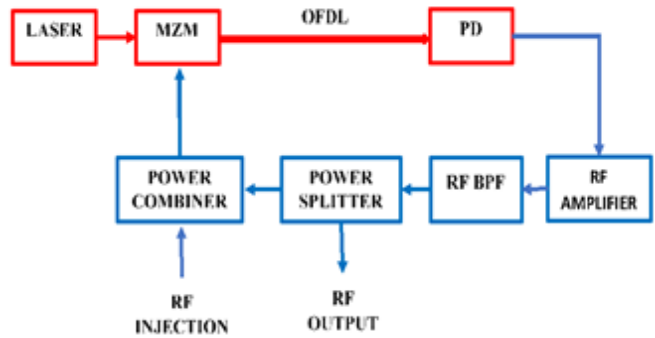

Fig. 1. Block diagram depicting a single-loop OEO under RF injection locking. OFDL: Optical fiber delay line, PD: Photo detector.

---

[1] The transfer function of the MZM is also non-linear and generates harmonics which are also ultimately dissipated by the RF BPF. The MZM could be used as the sole gain saturation mechanism. However, it is best to ensure that the saturated power delivered by the RF chain to the MZM is sufficient to drive the MZM over the interval between adjacent minimum and maximum transmission points and no further to maximize the modulation depth while avoiding an oscillator envelope instability [25] that occurs with higher drive levels.

where $u$ is the complex amplitude of the oscillation. The phase contributed by an intra-loop tuning element and other components is represented by $\phi$. The action of the RF BPF is represented as a convolution $\otimes$ by its impulse response $h$. The filter rejects the harmonics generated by the limiting of the sustaining amplifier. In addition, the transmission peak of its resonance aids the selection of the oscillating mode in a winner-takes-all gain competition provided by a gain control mechanism represented by the operator $K$, which is characterised by a large signal gain $k$ that decreases from the small signal gain in such away the amplitude of the fundamental harmonic is held substantially constant when the amplifier operates in saturation.

$$K(ae^{i\theta}) = kae^{i\theta} \sim a_0 e^{i\theta} \qquad (2)$$

The large signal gain is taken as real and positive as any associated phase including a sign is accounted for by $\phi$. The operator $D_\tau$ represents a delay:

$$D_{\tau_D} u(t) = u(t - \tau_D) \qquad (3)$$

where $\tau_D$ is the group delay of the loop provided substantially by the optical fibre delay line.

Equation (1) admits a family of freely oscillating solutions:

$$u(t) = a_p \exp(s_p t) \;\; ; \;\; s_p = \sigma_p + i\omega_p \qquad (4)$$

subject to the complex Barkhausen condition:

$$k\mathcal{H}(s_p)e^{i\phi} \exp(-s_p \tau_D) = 1$$
$$\Rightarrow \begin{cases} k|\mathcal{H}(s_p)| \exp(-\sigma_p \tau_D) = 1 \\ \omega_p \tau_D = 2p\pi + \phi + \arg(\mathcal{H}(s_p)) \quad p \in \mathbb{Z} \end{cases} \qquad (5)$$

where $\mathcal{H}$ is the Laplace transform of the impulse response $h$ of the RF filter. In the case of a single freely oscillating mode the envelope of the oscillation as time progresses grows exponentially if $k > 1$, decays exponentially if $k < 1$, and remains constant if $k = 1$. The gain control mechanism will consequently maintain $k = 1$ and a steady free oscillation will occur.

$$u(t) = a_p \exp(i\omega_p t + \theta_p) \qquad (6)$$

where $\theta_p$ is an arbitrary carrier phase at $t = 0$.

For $\phi = 0$ the number of cycles of the oscillating mode within the loop is an integer $p$. The change of frequency corresponding to a unit increment of the number of cycles within the loop defines the free-spectral range (FSR): the interval in frequency between allowed freely oscillating modes. The oscillation frequency may be tuned over a complete FSR by a suitable choice of $\phi \in (-\pi, \pi]$ that compensates for a fractional cycle contained within the delay line.

Multimode oscillation may occur following start-up. The transfer function of a RF BPF placed within the loop will modify the magnitude and phase of the loop gain as a function of the frequency of oscillation. A suitably peaked response will favor the mode that experiences the highest initial gain in a winner-takes-all gain competition leading to single-mode oscillation. Without loss of generality $u$ may be taken as the complex envelope of an oscillating carrier with a nominal frequency corresponding to the passband centre frequency of the RF BPF. The translation of the frequency variable to a baseband-offset frequency maps the impulse response of the BPF to the impulse response of a baseband equivalent low-pass filter (LPF).

The phase $\phi$ may be modelled as a quasi-static tuning phase $\phi_0$ additively perturbed by a phase-fluctuation $\delta\phi$. Provided the passband of the RF filter is sufficiently wide and either the frequency or the magnitude of the phase fluctuations is sufficiently small, the phase of the output signal of the RF filter is undistorted and equal to the phase of the input signal filtered by the baseband equivalent to the RF filter. This is the basis of the Leeson postulate that the phase is invariant to the exponent conjugation:

$$\arg(h \otimes \exp(i\theta)) = \arg(\exp(ih \otimes \theta)) \qquad (7)$$

The argument $\arg(z)$ of a complex number $z = x + iy$ may be evaluated using:

$$\arg(z) = \tan^{-1}(y/x) \tag{8}$$

where the arctangent should be interpreted over its range $(-\pi, \pi]$ with the signs of its numerator and denominator specifying the quadrant, c.f. the MATLAB function $\text{atan2}(y, x)$. The branch cut along the negative real axis is an artifact introduced by the polar co-ordinate system. If $z$ is a continuous function of time that avoids the origin, the branch cut can be eliminated by unwrapping the phase.

Applying the Leeson postulate to (1) yields a linear phase-only equation:

$$\theta_u = \phi + \left(h \otimes D_{\tau_D}\right)\theta_u \tag{9}$$

The fluctuation $\delta\theta_u$ of the carrier phase induced by the intra-loop phase fluctuation $\delta\phi$ satisfies the same equation by linearity. A Fourier transform then yields a relation between the associated spectra:

$$\widehat{\delta\theta_u}(\omega) = \frac{1}{(1-H(\omega)\exp(-i\omega\tau_D))}\widehat{\delta\phi}(\omega) \tag{10}$$

where $H$ denotes the Fourier transform of the impulse response $h$ of the RF filter. In the case of fluctuations described by a stationary stochastic process, it follows that:

$$S_{\delta\theta_u \delta\theta_u}(\omega) = \frac{1}{|1-H(\omega)\exp(-i\omega\tau_D)|^2} S_{\delta\phi\delta\phi}(\omega) \tag{11}$$

where $S_{\delta\theta_u \delta\theta_u}$ is the spectral density of the oscillator phase noise and $S_{\delta\phi\delta\phi}$ is the spectral density of the intra-loop phase fluctuations[2].

### III. SINGLE LOOP OEO SUBJECT TO RF INJECTION

In the presence of injection, the complex envelope $u$ representing the oscillation following the point of injection is the vector sum of the complex envelope $v$ representing oscillation prior to the point of injection and the complex envelope $w$ representing the injected carrier:

$$u = v + w \quad ; \quad \begin{cases} u = a_u \exp(i\theta_u) \\ v = a_v \exp(i\theta_v) \\ w = a_w \exp(i\theta_w) \end{cases} \tag{12}$$

In the case of a single-loop OEO it is convenient to apply the injection $w$ at the otherwise unused input port of the RF output coupler. This corresponds to $u$ representing the output of the coupler port driving the RF photonic link and $v$ representing the input to the coupler port driven by the RF chain. The complex envelope $v$ is a delayed replica of $u$ followed by saturated gain and therefore has a magnitude that may be assumed under certain restrictions to be substantially constant. The magnitude $w$ may also be taken as constant.

In general, the oscillation and the injection will interfere causing fluctuations of the magnitude of the vector sum $u$ but it will be assumed that its magnitude never falls below the level required to saturate the gain. This can be guaranteed for injection ratios outside a neighbourhood of unity that depends on the small signal gain and saturation power of the amplifier. Within this neighborhood it is necessary that the phase difference between $v$ and $w$ is sufficient to prevent the residual signal falling below the saturation threshold during an episode of destructive interference. With these caveats $a_v$ is taken as a constant. Now:

$$u = v + w \implies a_u e^{i(\theta_u - \theta_v)} = a_v + a_w e^{i(\theta_w - \theta_v)} \tag{13}$$

which may be expressed equivalently by:

$$re^{i\varphi} = 1 + \rho e^{i\theta} \tag{14}$$

where:

---

[2] In practice phase noise due to thermal noise following the RF filter provides an ultimate noise floor [26].

$$r = a_u/a_v \quad \varphi = \theta_u - \theta_v \quad \rho = a_w/a_v \quad \theta = \theta_w - \theta_v \tag{15}$$

The injection therefore induces a change in the relative magnitude $r$ and phase $\varphi$ of the superposition that is controlled by the phase difference $\theta$ between injected carrier and the oscillation and the injection ratio $\rho$. The phase shift $\varphi$ is responsible for tuning the oscillator via the Barkhausen phase condition for oscillation and the relative magnitude $r$ is responsible via the gain control mechanism for locking.

Equating the real and imaginary parts of (14):

$$\begin{aligned} r \cos(\varphi) &= 1 + \rho \cos(\theta) \\ r \sin(\varphi) &= \rho \sin(\theta) \end{aligned} \tag{16}$$

and eliminating $r$ yields:

$$\varphi = \tan^{-1}\left(\frac{\rho \sin(\theta)}{1 + \rho \cos(\theta)}\right) \quad ; \quad \theta = \theta_w - \theta_v \tag{17}$$

where the arctangent is to be interpreted in the four-quadrant sense. Alternatively, eliminating $\theta$ from (16) yields the quadratic equation:

$$r^2 - 2\cos(\varphi)\, r + (1 - \rho^2) = 0 \tag{18}$$

A real solution that is positive for all $\rho$ is given by:

$$r = \cos(\varphi) + \sqrt{\rho^2 - \sin^2(\varphi)} \tag{19}$$

subject to the condition:

$$\rho^2 \geq \sin^2(\varphi) \tag{20}$$

The evolution of the phase of the oscillation is described by a system of three equations corresponding to: the Leeson model of the RF filter following the delay line (21); the phase modulation due to the injected carrier (22); and the Barkhausen phase condition for oscillation (23):

$$\theta_v = (h \otimes D_{\tau_D})\theta_u \tag{21}$$

$$\varphi = \tan^{-1}\left(\frac{\rho \sin(\theta_w - \theta_v)}{1 + \rho \cos(\theta_w - \theta_v)}\right) \tag{22}$$

$$\theta_u = \varphi + \phi + \theta_v \quad mod\ 2\pi \tag{23}$$

In the special case of a single pole approximation of the low pass baseband equivalent model of a band-pass RF filter, (21) may be written:

$$\tau_R \frac{d\theta_v}{dt} + \theta_v = D_{\tau_D}\theta_u \tag{24}$$

where $\tau_R$ is a time constant characterizing the bandwidth of the filter[3]. The system may be combined into a single recursive equation which forms the principal result of this paper.

$$(I - h \otimes D_{\tau_D})\theta_v = (h \otimes D_{\tau_D})\left[\phi + \tan^{-1}\left(\frac{\rho \sin(\theta_w - \theta_v)}{1 + \rho \cos(\theta_w - \theta_v)}\right)\right] \tag{25}$$

Consider the special case where the content of the square bracket is a Heaviside step of height $\phi$. It follows from the identity:

$$(I - h \otimes D_{\tau_D})^{-1}(h \otimes D_{\tau_D}) = \sum_{n=1,\infty}(h \otimes)^n D_{n\tau_D} \tag{26}$$

that the oscillator accumulates the phase step into an infinite staircase function. The step placed upon the staircase at time $t = n\tau_D$ is smoothed by $n$-fold convolution by the RF filter impulse response. The smoothing contributed by each convolution is slight, but the leading edge of the step broadens indefinitely. Consequently, the staircase evolves into a linear ramp:

$$\theta_u = \omega t \quad ; \quad \omega \tau = \phi \mod 2\pi \quad ; \quad \tau = \tau_D + \tau_R \tag{27}$$

which may be verified by a straightforward direct computation.

---

[3]The time constant $\tau_R$ is the group delay of the RF filter at the passband centre frequency.

The slope of the linear ramp $\omega$ provides the frequency of the equilibrium oscillation[4] relative to the nominal reference frequency. If the ramp is subtracted from the smoothed staircase the result is an almost periodic function with a period asymptotic to the total round-trip group delay $\tau$. The components of the Fourier series that represents a local period of the function generate sidemodes. The RF filter lightly suppresses these sidemodes on each round trip causing the evolution of the phase into a ramp. This evolution is grounded on the ultimate selection of a single mode of oscillation by the winner-takes-all gain competition process.

The oscillator converts $\varphi$ via the Barkhausen condition (23) into a change of oscillation frequency[5]; the time delay oscillator behaves as a phase-controlled oscillator. For small injection $\rho \ll 1$, (22) simplifies to:

$$\varphi = \rho \sin(\theta) \quad ; \quad \theta = \theta_w - \theta_v \tag{28}$$

which is identical to the phase shift provided by an intra-loop phase shifter driven, via a loop amplifier with dimensionless gain $\rho$, by an analog phase sensitive detector[6] with a phase difference $\theta = \theta_w - \theta_v$ between its reference $\theta_w$ and signal $\theta_v$ inputs. The oscillator under injection is analogous to a proportionally controlled PLL. The analogy only fails to be an equivalence because of the difference for large injection between (22) and (28). Nevertheless, the behavior of the oscillator under injection remains qualitatively similar.

## IV. COMPARISON WITH CLASSICAL INJECTION LOCKING THEORY

Equation (25) encompasses the evolution equations of classical injection locking theory in two limiting cases corresponding to either $\tau_D \ll \tau_R$ or the converse $\tau_R \ll \tau_D$.

In the first case the round-trip time is dominated by the group delay of the RF resonator $\tau = \tau_R$ so that $D_{\tau_D} \to I$ and one may assume all variables vary slowly over the round-trip time[7]. Convolution by the RF resonator impulse response then approaches the identity and:

$$\left(I - h \otimes D_{\tau_D}\right)\theta_u \to \theta_u - \theta_v = \tau_R \frac{d\theta_v}{dt} \tag{29}$$

Consequently, (25) reduces to:

$$\tau_R \frac{d\theta_v}{dt} = \phi + \tan^{-1}\left(\frac{\rho \sin(\theta_w - \theta_v)}{1 + \rho \cos(\theta_w - \theta_v)}\right) \tag{30}$$

It is convenient to rewrite (30) in terms of the phase difference $\theta = \theta_w - \theta_v$ between the injected carrier and the oscillation. In the special case of the injection of a pure carrier of frequency $\omega_I$ and noting $\phi = \omega_O \tau_R$ one obtains:

$$\tau_R \frac{d\theta}{dt} = (\omega_I - \omega_O)\tau_R - \tan^{-1}\left(\frac{\rho \sin(\theta)}{1 + \rho \cos(\theta)}\right) \tag{31}$$

In the second case the round-trip time is dominated by the delay line $\tau = \tau_D$ and one may assume all variables vary slowly over a time scale of $\tau_R$. Convolution by the RF resonator impulse response approaches the identity and (25) reduces to:

---

[4]The Leeson postulate is responsible for the linear approximation of the arctangent function in the exact expression $\omega \tau_D + \tan^{-1}(\omega \tau_R) = \phi \mod 2\pi$.

[5]The oscillator phase increments by $\varphi$ on each round trip of the RF oscillation; the cumulative sum approximates an integral on time scales large compared to the round-trip time.

[6]An analog phase sensitive detector consists of balanced mixer (4-quadrant multiplier) with the oscillator output applied to one input port and the injected carrier applied to the other input port with the component at the sum frequency removed from the output.

[7]This is the case if the phase increment per round trip is small, which is the case for weak injection.

$$(I - D_{\tau_D})\theta_v = D_{\tau_D}\left[\phi + \tan^{-1}\left(\frac{\rho \sin(\theta_w - \theta_v)}{1 + \rho \cos(\theta_w - \theta_v)}\right)\right] \tag{32}$$

It is convenient to rewrite (32) in terms of the phase difference $\theta = \theta_w - \theta_v$ between the injected carrier and the oscillation. In the special case of the injection of a pure carrier of frequency $\omega_I$ and a constant tuning phase $\phi = \omega_O \tau_D$ one obtains:

$$(I - D_{\tau_D})\theta = (\omega_I - \omega_O)\tau_D - D_{\tau_D}\tan^{-1}\left(\frac{\rho \sin(\theta)}{1 + \rho \cos(\theta)}\right) \tag{33}$$

The action of the RF filter plays an important implicit role in ensuring the validity of the assumptions that lead to (32) and (33). It is now further assumed that the RF filter in conjunction with the gain control mechanism is successful in establishing single mode oscillation prior to the onset of injection and consequently that all variables vary slowly over a round trip time[7] which is determined substantially by the delay line $\tau = \tau_D$. The delay operator $D_{\tau_D}$ approaches the identity and:

$$(I - D_{\tau_D})\theta = \int_{t-\tau_D}^{t} \frac{d\theta}{dt}(t')dt' \approx \tau_D \frac{d\theta}{dt} \tag{34}$$

by the mean value theorem. Consequently (33) reduces to:

$$\tau_D \frac{d\theta}{dt} = (\omega_I - \omega_O)\tau_D - \tan^{-1}\left(\frac{\rho \sin(\theta)}{1 + \rho \cos(\theta)}\right) \tag{35}$$

The only difference between (31) and (35) is a change of the time constant parameter which is $\tau_R$ when the RF resonator is dominant and $\tau_D$ when the delay line is dominant. Equations (31) and (35) may be recognized as a variant of the Paciorek equation [19]. Paciorek and Adler treat the RF resonator in the quasi-stationary approximation, which replaces the Leeson model of the combined RF filter and delay line (16) by:

$$\theta_v = \arg\left(H\left(\frac{d\theta_v}{dt}\right)\right) + D_{\tau_D}\theta_u \tag{36}$$

Substitution of (36) into the Barkhausen phase condition (23) yields:

$$(I - D_{\tau_D})\theta_v = D_{\tau_D}(\varphi + \phi) + \arg\left(H\left(\frac{d\theta_v}{dt}\right)\right) \;\; mod \; 2\pi \tag{37}$$

which for an equilibrium oscillation with frequency $\omega_O$ reduces to:

$$\omega_O \tau_D = \phi + \arg(H(\omega_O)) + \varphi \quad mod \; 2\pi \tag{38}$$

Equation (38) reproduces the oscillation frequencies of the free ($\varphi = 0$) oscillator predicted by (5). Paciorek and Adler do not consider a non-zero delay $\tau_D$ or tuning phase $\phi$; the free oscillation frequency is determined solely by the RF resonator:

$$\arg(H(\omega_O)) = 0 \tag{39}$$

Hence, for an injected carrier of frequency $\omega_I$:

$$\arg(H(\omega)) + \varphi = 0 \; ; \; \omega = (\omega_I - \omega_O) - \frac{d\theta}{dt} \tag{40}$$

In the special case of a single pole baseband-equivalent model of a band-pass RF filter:

$$-\tan^{-1}(\omega \tau_R) = \varphi = \tan^{-1}\left(\frac{\rho \sin(\theta)}{1 + \rho \cos(\theta)}\right) \tag{41}$$

Consequently, one arrives at the original Paciorek equation:

$$\tau_R \frac{d\theta}{dt} = (\omega_I - \omega_O)\tau_R - \frac{\rho \sin(\theta)}{1 + \rho \cos(\theta)} \tag{42}$$

after elimination of the arctangent functions. In the weak injection limit $\rho \to 0$ both the Paciorek equation and its variant (42), (31) reduce to the Adler equation [18][8].

$$\tau \frac{d\theta}{dt} = (\omega_I - \omega_O)\tau - \rho \sin(\theta) \tag{43}$$

A. *Equilibrium Solutions: Existence*

An equilibrium solution to (35) exists if:

$$(\omega_I - \omega_O)\tau_D = \tan^{-1}\left(\frac{\rho \sin(\theta_\infty)}{1+\rho \cos(\theta_\infty)}\right) \tag{44}$$

has a solution for $\theta_\infty$. A solution always exists if $\rho \geq 1$; two solutions exist if $\rho < 1$ provided the detuning falls within the locking range:

$$(\omega_I - \omega_O)\tau_D \in [-\sin^{-1}(\rho), \sin^{-1}(\rho)] \tag{45}$$

and otherwise, there is no equilibrium solution. Equation (45) is equivalent to condition (20).

B. *Equilibrium Solutions: Stability*

Linearization of (35) yields:

$$\tau_D \frac{d\delta\theta}{dt} + \left\{\frac{\rho \cos(\theta_\infty)+\rho^2}{1+2\rho \cos(\theta_\infty)+\rho^2}\right\} \delta\theta = 0 \tag{46}$$

The denominator of the term in braces is always positive consequently a necessary condition[9] for stability is:

$$\cos(\theta_\infty) + \rho > 0$$
$$\Rightarrow \begin{cases} \theta_\infty \in [-\pi/2 - \sin^{-1}(\rho), \pi/2 + \sin^{-1}(\rho)] & ; \rho \leq 1 \\ \theta_\infty \in [-\pi, \pi] & ; \rho > 1 \end{cases} \tag{47}$$

Equations (45) and (47) are different statements of the same geometry. Provided the detuning falls within the range specified by (45) of the oscillating mode, the equilibrium solution specified by (44) is attractive.

An injection level $\rho = 1$ is sufficient to achieve locking over one half of an FSR. The phase-only model places no restriction on the locking range for $\rho > 1$ and locking over a full free spectral range appears possible. The predicted locking range is mathematically correct for the phase-only model. However, magnitude perturbations cannot be ignored at high injection levels. The gain control mechanism then plays a critical role and recourse to an envelope model is necessary. The induced relative magnitude (19) is unity for no injection ($\rho = 0$, $\varphi = 0$) and exceeds unity for injection with no detuning. Via the gain control mechanism, the increased relative magnitude $r > 1$ during episodes of constructive interference between the injected carrier and the oscillation reduces the gain of the sustaining amplifier below the threshold for free oscillation. It is then the injected carrier that drives the oscillator. if there are sufficiently sustained episodes of constructive interference, the injection ultimately captures the oscillator.

Setting $r = 1$, Equation (19) simplifies to:

$$4 \sin^2(\varphi/2) = \rho^2 \tag{48}$$

It follows that the envelope model locking range is:

$$(\omega_I - \omega_O)\tau_D \in [-2\sin^{-1}(\rho/2), 2\sin^{-1}(\rho/2)] \tag{49}$$

An injection level of $\rho = 2$ is sufficient to achieve locking over a full FSR. For small injection:

$$\sin^{-1}(\rho) \to 2\sin^{-1}(\rho/2) \tag{50}$$

---

[8] Paciorek and Adler make a choice of $\theta$ that differs in sign from the choice herein which has the effect of exchanging $\omega_I$ and $\omega_O$ in (31), (33), (35), & (38).
[9] The injection ratio is positive.

to high accuracy. Implicit injection levels up to unity can be found in some dual loop or dual oscillators but the low injection regime is of interest in most applications. The locking range predicted by phase only models (this work, Paciorek, Adler) substantially agree in the weak injection regime of most common interest. Nevertheless, implicit injection levels up to unity occur in some self-injection or cross-injection oscillator topologies. Whereas the phase-only models correctly describe the tuning of the oscillator by the phase shift induced by injection, it is the gain control mechanism that is responsible for the locking phenomenon.

The asymptotic phase shift between the locked oscillation and the injected carrier is zero only for zero detuning and for weak injection approaches $\pm \pi/2$ at the edge of the locking range. Unlocked, the free-oscillation and the injected carrier will co-exist within the oscillator and interfere; alternating between episodes of essentially destructive and constructive interference. An episode of constructive interference is sufficient for the stability of an equilibrium solution. The injected signal will therefore capture the oscillator provided the detuning is within the locking range of the oscillating mode. Since an episode of constructive interference is sufficient for an attractive equilibrium solution the restriction that the magnitude of the vector sum should not fall below the saturation threshold is of little consequence and any limitation on the injection ratio may be lifted at least in respect of the locking dynamics.

The elimination of the arctangent functions in the Paciorek equation for a resonator dominated oscillator results in the prediction of a locking range:

$$(\omega_I - \omega_O)\tau_R \in \left[-\frac{\rho}{\sqrt{1-\rho^2}}, \frac{\rho}{\sqrt{1-\rho^2}}\right] \tag{51}$$

which extends towards infinity as the injection ratio approaches unity. The question arises whether the elimination of the arctangent functions by use of the quasi-stationary approximation is meaningful outside the linear region of the respective arctangent functions. It is certainly the case that the arctangent appearing in the expression for the injection phase shift is correct for any injection level but one cannot be so certain of the accuracy of the phase-only transfer function of the RF filter at instantaneous frequencies beyond its linear region where RF filter becomes increasingly attenuating. These considerations impact only resonator dominated oscillators. The Leeson model provides an adequate description of the RF filter for delay dominated oscillators.

This work applies the Leeson postulate which has the advantage of providing a linear RF phase-only transfer function which is valid provided the slow frequency fluctuations and the fast phase fluctuations are small. Moreover, it provides an exact result for a time delay. Consequently, the arctangent of the injection phase-shift term remains in the resonator dominated case (30) and its retention is unavoidable in the delay dominated case (32).

## V. LINEAR MODEL AND PHASE NOISE

The linearization of (25) about a locked state is given by:

$$\delta\theta_v = (h \otimes D_{\tau_D})(\delta\phi + \eta\delta\theta_w + (1-\eta)\delta\theta_v) \tag{52}$$

where $\eta$ is the linearized proportional gain given by:

$$\eta = \frac{\rho \cos(\theta_\infty) + \rho^2}{1 + 2\rho \cos(\theta_\infty) + \rho^2} \tag{53}$$

where $\theta_\infty$ is the asymptotic phase difference between the oscillator and the injected carrier which satisfies (44). The proportional gain attains a maximum value:

$$\eta = \frac{\rho}{1+\rho} \quad ; \quad \theta_\infty = 0 \tag{54}$$

corresponding to zero detuning. The gain reduces with decreasing $\cos(\theta_\infty)$ and reaches zero:

$$\eta = 0 \quad ; \quad \cos(\theta_\infty) = -\rho \tag{55}$$

at the limits of the phase-only locking range. The gain is negative with further decrease of $\cos(\theta_\infty)$ indicating an unstable equilibrium. The gain attains a most negative value:

$$\eta = \frac{-\rho}{1-\rho} \quad ; \quad \theta_\infty = \pm\pi \tag{56}$$

corresponding to detuning to the mid-point in frequency between adjacent modes.

Consequently zero-detuning is most advantageous from the perspective of proportional control. Moreover, it is found that the simple linearized model of (52) with $\eta = \rho/(1+\rho)$ is an accurate approximation to the full nonlinear model in a substantial neighborhood of $\theta_\infty = 0$.

In the analysis of injection locking, it is natural to designate $v$ as the output of the oscillator. The response of the oscillator to intra-loop and injected carrier phase fluctuations is then described by (52). However, in practice it is more convenient to use the spare output port of the coupler used to inject $w$ which designates $u$ as the output. The appropriate equation is:

$$\delta\theta_u = \delta\phi + \eta\delta\theta_w + (1-\eta)\left(h \otimes D_{\tau_D}\right) \tag{57}$$

Taking the Fourier transform of (57) and rearranging terms, the phase noise spectrum of the oscillator is given by:

$$\widehat{\delta\theta_u}(\omega) = \frac{1}{1-(1-\eta)H(\omega)\exp(-i\omega\tau_D)}\left(\widehat{\delta\phi}(\omega) + \eta\widehat{\delta\theta_w}(\omega)\right) \tag{58}$$

which is agreement with the result for the free oscillator (10) in the absence of injection ($\eta = 0$). For fluctuations $\delta\phi$, $\delta\theta_w$ described by uncorrelated stationary stochastic processes, the spectral density of the oscillator phase noise is:

$$S_{\delta\theta_u\delta\theta_u}(\omega) = \frac{1}{|1-(1-\eta)H(\omega)\exp(-i\omega\tau_D)|^2}\left(S_{\phi\phi}(\omega) + \eta^2 S_{\delta\theta_w\delta\theta_w}(\omega)\right) \tag{59}$$

where $S_{\phi\phi}, S_{\delta\theta_w\delta\theta_w}$ are the spectral density of the phase fluctuations internal to the oscillator and externally injected into the oscillator, respectively.

It is notable and intuitive that the phase noise of the injected carrier weighted by $\eta$ contributes additively to the intra-loop phase fluctuations of the oscillator. The denominator of the first factor of (58) is responsible for introducing spurious resonances in the phase-noise spectrum. These spurious resonances are substantially suppressed as the level of injection is increased relative to the same term without injection ($\eta = 0$). This is because the third term on the right-hand side of (58) reduces the influence of the prior state of the oscillator (memory) and increases the influence of the injection (innovation). Indeed, the pole at zero frequency and other resonances characteristic of the integrating action of a free oscillator are replaced in a neighborhood of each resonance by a BPF response with high finesse $\mathcal{F}$ for small injection:

$$\frac{\eta}{1-(1-\eta)\exp(-i\omega\tau_D)} \sim \frac{1}{1+i\omega((1-\eta)/\eta)\tau_D}; i\omega\tau_D \to 0$$
$$\Rightarrow \mathcal{F} = \pi/\eta \quad ; \quad \eta \ll 1 \tag{60}$$

Consequently, decreasing the level of injection is effective at narrowband filtering of the phase noise of the external injected carrier about every resonance, but it is of no aid in the reduction of the intra-loop phase fluctuations as these are not weighted by $\eta$. This may be understood as the accumulation of the quasi-static phase fluctuations on every round-trip. Conversely, increasing the level of injection is effective at suppressing spurious resonances but increases the phase noise contributed by the injected carrier.

## VI. EXPERIMENTAL RESULTS

Experiments based on the schematic in Fig. 2 have been performed to validate the analytical results. A wavelength-division-multiplexed (WDM) source laser provides the RF photonic link with an optical power of 13 dBm at a vacuum wavelength of 1.55 μm. A low-phase noise RF source provides the injected carrier. The three amplifiers in the RF chain are all ADL5602. A surface-acoustic-wave (SAW) resonator with a bandwidth of 270 kHz is used as the RF mode selection filter (BPF). The fibre coil has a length of 5.050 km. The natural frequency of the OEO is 1.090115600 GHz. This is also the frequency of the RF injection for zero detuning.

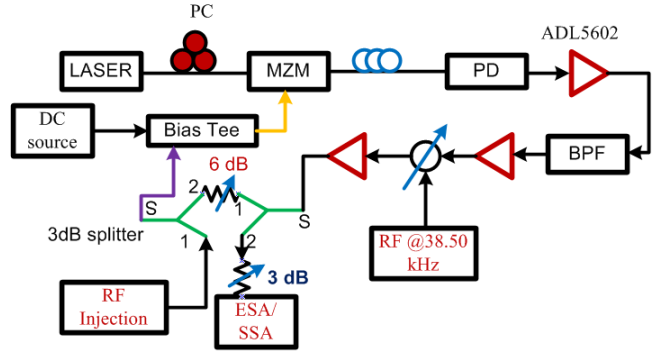

Fig. 2. Schematic diagram of the experimental arrangement. PC: polarization controller, ESA/SSA electronic spectrum analyser / signal source analyser.

A low frequency RF signal generator provides an option to add an artificial component to the natural intra-loop phase fluctuations to facilitate measurements. The application of a 100-mV peak to peak sinusoid at a frequency 38.73 kHz to the tuning phase shifter (shown in the schematic as a circle crossed by an arrow placed between two RF amplifiers) generates weak intra-loop phase modulation (1.2°) resonant with the 1$^{st}$ sidemode which increases the sidemode levels.

The small network of splitters and attenuators provides an injection port and a monitoring port. The latter permits the power of the fundamental mode and sidemodes and the phase noise spectral density to be measured by the ESA/SSA. The losses, attenuation, and splitting ratio of the individual components including the cables are fully characterized in a separate experiment to enable the injection ratio to be calculated from the power output of the RF injection source and the input power of the fundamental mode observed at the ESA/SSA.

Fig. 3 shows a comparison of the locking range as a function of injection ratio obtained from the experiment with the analytical result of (45) (phase model), and (49) (saturating gain model). The phase-only model predicts the observed locking range well only for injection ratios $\rho < \sim 0.7$ and fails completely to capture the observed behaviour for injection ratios $\rho > 1$, as described in Section IV. The saturating gain model provides an excellent prediction of the observed behavior over the complete range of injection ratios $\rho < 2$.

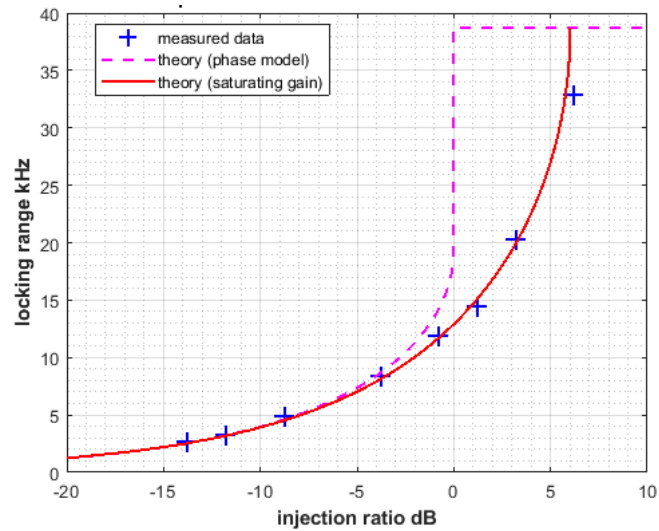

Fig. 3. Comparison of theoretically calculated and measured locking ranges of injection locked OEO as a function of injection ratio.

Fig. 4 shows the measured phase noise spectral densities of the RF injection source (cyan curve); the free running (no injection or modulation) OEO (blue curve); the free running (no injection) OEO with intra loop phase modulation; and the injection locked OEO with intra loop phase modulation. The phase noise spectral density of the free running OEO exceeds that of the RF source at frequencies below 1 kHz. Above 1 kHz the phase noise of the injection source and the phase noise floor of the OEO are comparable. The phase noise spectrum of the OEO contains clear evidence of spurious sidemode resonances above the noise floor with the expected FSR ~39 kHz. The first sidemode resonance has a peak ~ 20 dB above the noise floor[10]. These sidemodes are well suppressed by the narrow bandwidth ~270 kHz of the 1 GHz RF filter relative to the sidemode FSR. Weak intra-loop phase modulation is applied via the tuning phase shifter at a frequency tuned to the first side mode resonance. The magnitude of the sidemodes is thereby enhanced, as shown in Fig. 4 (red curve), facilitating the measurement of their suppression by RF injection at the frequency of the fundamental mode. With increasing injection ratio, the low frequency phase noise of the OEO more closely follows the phase noise of the RF injection source within the locking-range and the sidemodes are increasingly suppressed (Fig. 6, magenta curve). A phase noise of -120 dBc/Hz@ 1 kHz offset is attained for a RF injection ratio of 1.21 dB.

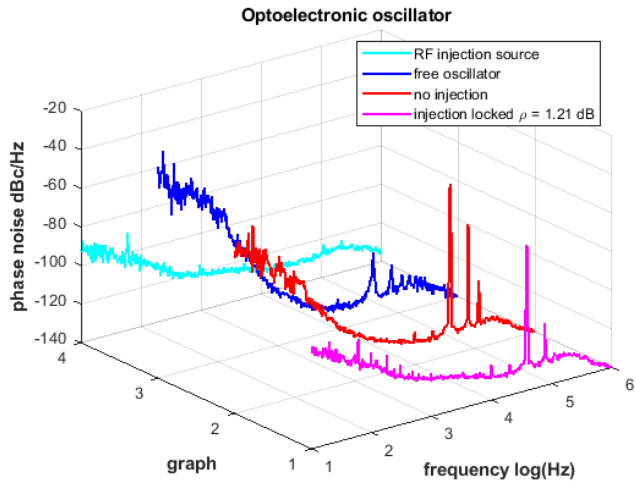

Fig. 4. Measured phase noise of RF injection source, free running OEO (without injection and without any modulation), free running OEO with intra loop phase modulation (without injection) and injection locked OEO with intra loop phase modulation.

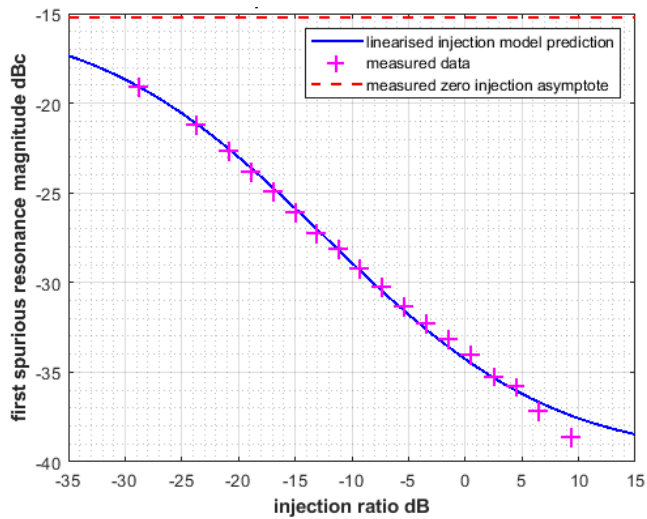

Fig. 5. Comparison of theoretically calculated and measured first spurious resonance level as a function of injection ratio.

Fig. 5 shows the dependence of the first spurious resonance magnitude on the injection ratio of the injection locked OEO. Although a straight line fits the measured log-log data for non-zero injection reasonably well, it does not fit the zero-injection measurement which sets the -15.22 dB asymptote. Moreover, the measured data suggests an inverted 'S' shape over injection levels below 0dB which is captured by the theoretical prediction. The theoretical prediction makes use of the linearized model of injection locking that is not valid for injection levels exceeding 0dB. The measured data hints at a change of slope at 0 dB injection suggesting a qualitative change of behaviour in the large injection regime. The best fit is consistent with the asymptote for a detuning of 8.5 kHz of the passband centre of the RF filter from the natural frequency of the oscillator. A small detuning is plausible given that the frequency spacing of the modes ~ 39 kHz is a significant proportion of the bandwidth ~ 270 kHz of the RF filter.

---

[10] The spectral densities underestimate the sidemode level as their linewidths are small compared to the resolution bandwidth of the analyzer.

## VII. CONCLUSIONS

This paper has introduced a delay-differential equation description of injection locking of time delay oscillators. The formulation captures the essential features of the dynamics and phase noise of the time delay oscillator. Unlike prior treatments that assume a single mode oscillator, it successfully predicts the effect of injection locking on the spurious resonances in the phase noise spectrum corresponding to the sidemodes of a time delay oscillator. Moreover, it is not restricted to weak injection. A single loop OEO injection locked by an external source is used as an example to elucidate the fundamental theory and to provide experimental validation of findings such as locking range and sidemode suppression that differ or are absent in prior treatments. It is straightforward to extend the formulation to oscillator topologies of greater complexity such as self- or cross- injection locking of single-loop, multi-loop, or multi-oscillator architectures.